\documentclass[%
 reprint,
 amsmath,amssymb,
 aps, pra,
floatfix,
]{revtex4-2}

\usepackage{graphicx}
\usepackage{dcolumn}
\usepackage{bm}
\usepackage[mathlines]{lineno}
\usepackage{mathtools}
\usepackage{multirow}
\usepackage{adjustbox}
\usepackage{xcolor} 
\usepackage{makecell}

\begin{document}
\title{Efficient ultrafast homodyne detection of quantum light}
\author{Young-Do Yoon, Chan Roh, Geunhee Gwak}
\author{Young-Sik Ra}
\email{youngsikra@gmail.com}
\affiliation{Department of Physics, Korea Advanced Institute of Science and Technology (KAIST), Daejeon, 34141, Korea}
\date{\today}

\begin{abstract}
Ultrafast continuous-variable quantum states offer new opportunities for advanced quantum technologies, but efficient homodyne detection of these states remains challenging. Here, we present a method for efficient ultrafast homodyne detection by exploiting temporal correlations in detector signals. By optimizing the temporal weight used to extract quadrature outcomes, we achieve a substantial increase in the signal-to-noise ratio of ultrafast homodyne detection, thereby improving the detection efficiency. We analyze the autocorrelations of shot noise and electronic noise and determine the optimal weight by solving a generalized Rayleigh quotient problem. The optimal weight enhances the squeezing and anti-squeezing levels observed experimentally. These results highlight the importance of optimized signal processing for efficient quantum measurements.
\end{abstract}

\maketitle

\section{Introduction}

Continuous-variable quantum optical systems offer a promising route toward scalable quantum technologies~\cite{Larsen2019, Asavanant2019, Guo2020, Kovalenko2021, Deng2023, AghaeeRad2025, Jia2025, Liu2025}. Although these systems have primarily been developed in the continuous-wave regime~\cite{Andersen2016}, their extension to the ultrafast regime has recently attracted considerable attention~\cite{Casacio2021, Nehra2022, Rasputnyi2024, Herman2025, Sennary2025, Cimini2026}. Ultrafast light pulses, characterized by broad spectral distributions, can support multiple optical modes for generating large-scale entangled states~\cite{Ra2020, Kouadou2023, Roh2025, Kalash2025, Gwak2026, Roh2026}. In addition, ultrafast quantum states can occupy localized temporal modes, a desirable feature for photon-number-resolving detection~\cite{Bimbard2010, Deng2023, Endo2023, Simon2024}. The high optical energy contained in each pulse further facilitates efficient nonlinear interactions, enabling single-pass generation of squeezed light~\cite{Casacio2021, Deng2023, Barakat2025, Kalash2025, Cimini2026}.  Ultrafast systems also offer advantages for specific quantum applications, such as precise timing measurements in quantum metrology~\cite{Lamine2008, Wang2018} and high information-processing rates in quantum communication and computation~\cite{Kovalenko2021, Sennary2025, Endo2023}.

Homodyne detection is a standard method for measuring continuous-variable quantum systems, providing quadrature outcomes of nonclassical light fields. In the continuous-wave regime, homodyne detection is a well-established technique~\cite{Bachor2019}, in which the detector can directly capture slowly varying optical signals. In the ultrafast regime, however, the detector response is too slow to resolve the rapid variation of the optical field; to achieve ultrafast homodyne detection, interference with ultrafast local-oscillator pulses is used~\cite{Kumar2012, Cooper2013}. This produces pulsed electrical signals, in contrast to the nearly uniform signals in continuous-wave homodyne detection. The large peaks of these pulsed signals, however, can readily saturate the detector response, limiting the allowable local-oscillator power in practice~\cite{Kumar2012, Cooper2013}. Broadening the detector response to reduce the peak height is not a viable solution~\cite{Masalov2017}, because it can hinder the discrimination of adjacent pulses~\cite{Kumar2012, Cooper2013}. Consequently, the signal-to-noise ratio (SNR) of ultrafast homodyne detection has remained limited~\cite{Hansen2001, Zavatta2006, Okubo2008, Kumar2012, Cooper2013, Du2018, Kouadou2026}.

This limitation is important because the SNR is directly connected to the detection efficiency in homodyne detection. The SNR is defined as the ratio of the shot-noise variance to the electronic-noise variance, and a limited SNR is equivalent to an optical loss that reduces the detection efficiency~\cite{Appel2007}.  This effective loss degrades nonclassical features observed in quantum states, such as squeezing and Wigner-function negativity, with stronger effects on highly nonclassical states~\cite{Endo2023, Vahlbruch2016}. Because such nonclassical features are central to many protocols in quantum computing~\cite{Larsen2019, Asavanant2019, Mari2012}, quantum communication~\cite{Kovalenko2021, Sennary2025}, and quantum metrology~\cite{Guo2020, Lamine2008, Aasi2013}, achieving a high SNR is essential for efficient detection of quantum states.

In this work, we devise a method to enhance the SNR of ultrafast homodyne detection by exploiting temporal correlations in the detector signals. We observe that the pulsed detector signals in ultrafast homodyne detection exhibit strong temporal correlations that are absent in the electronic noise. Consequently, an optimized temporal weight can efficiently extract quadrature outcomes from the signals while suppressing the contribution of electronic noise. This optimization leads to a substantial increase in the SNR. We show that the optimization can be formulated as a generalized Rayleigh quotient problem, which also appears in quantum parameter-estimation theory~\cite{Gessner2019, Go2026}. We then present a systematic method for analyzing temporal correlations in detector signals and determining the optimal weight for efficient ultrafast homodyne detection. Existing methods for extracting quadrature outcomes can also be described within the same framework: the area-integration method~\cite{Zavatta2006, Okubo2008, Kumar2012, Cooper2013, Kouadou2026} and the peak-value detection method~\cite{Hansen2001, Du2018} correspond to using constant and peak-shaped weights, respectively. Experimentally, we find that the optimal weight differs from the conventional weights and improves the homodyne detection efficiency. We further show that this improvement directly enhances the observed squeezing and anti-squeezing levels in measurements of ultrafast squeezed vacuum.

\section{Theory of ultrafast homodyne detection}

\begin{figure}[t]
	\centering
	\includegraphics[width=8.5cm]{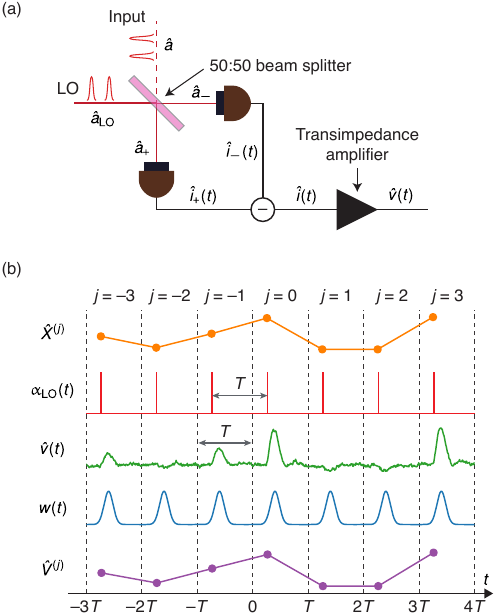}
	\caption{Ultrafast homodyne detection. (a) An input field interferes with the local oscillator (LO) at the 50:50 beam splitter, and two photodiodes detect the output beams and convert them into photocurrents $\hat{i}_{\pm}(t)$. The photocurrent difference $\hat{i}(t)=\hat{i}_{+}(t) - \hat{i}_{-}(t)$ is converted into a voltage signal $\hat{v}(t)$ by the transimpedance amplifier. (b) A homodyne detection signal and its processing for obtaining quadrature outcomes. $\hat{X}^{(j)}$ shows the quadrature outcome for the $j$-th pulse, and $\alpha_{\text{LO}}(t)$ depicts the time-varying amplitude of the LO, which is formed by a train of ultrafast pulses with repetition period $T$. The resulting voltage signal $\hat{v}(t)$ is weighted by the weight function $w(t)$ and integrated, yielding the processed signal $\hat{V}^{(j)}$.
    }
	 \label{fig:BHD}
\end{figure}

Figure \ref{fig:BHD}(a) illustrates a homodyne detection scheme for measuring the quadratures of an input field. The input field interferes with the local oscillator (LO) at a 50:50 beam splitter, and the resulting output fields are detected by two photodiodes. The associated annihilation operators are denoted by $\hat{a}(t)$ for the input, $\hat{a}_{\text{LO}} (t)$ for the LO, and $\hat{a}_{\pm}(t)$ for the two outputs. The input field operator satisfies the commutation relations $[\hat{a}(t), \hat{a}^{\dagger}(t^{\prime})] = \delta (t - t^{\prime})$ and $[\hat{a}(t), \hat{a}(t^{\prime})] = 0$, and $\hat{a}_{\text{LO}} (t)$ can be replaced by a classical amplitude $\alpha_{\text{LO}} (t)$ since the LO is a strong laser field. The output field operators $\hat{a}_{\pm}(t)$ can be expressed in terms of the input operator and the LO amplitude as follows:
\begin{equation}
	\hat{a}_{\pm}(t) = \frac{1}{\sqrt{2}} \left( \hat{a}(t) \pm \alpha_{\text{LO}} (t)\right).
\end{equation}
The photodiodes convert the photon fluxes $\hat{n}_{\pm}(t)=\hat{a}_{\pm}^{\dagger}(t)\hat{a}_{\pm}(t)$ into photocurrents $\hat{i}_{\pm}(t) = e\hat{n}_{\pm}(t)$, where $e$ denotes the electron charge.
These photocurrents are then subtracted,
\begin{equation}
	\hat{i}(t) = \hat{i}_{+}(t) - \hat{i}_{-}(t) = e |\alpha_{\text{LO}} (t)| \hat{x}_{\theta}(t).
	\label{eq:ideal}
\end{equation}
Here the LO amplitude $\alpha_{\text{LO}} (t)$ is written as $|\alpha_{\text{LO}} (t)| e^{i\theta}$, where $\theta$ is the relative phase between the LO and the signal, and $\hat{x}_{\theta} (t) = \hat{a}(t) e^{-i\theta}+ \hat{a}^{\dagger}(t) e^{i\theta}$ represents the input quadrature operator at phase $\theta$. The dark current of the photodiodes, being typically negligible, is omitted. At the final step, a transimpedance amplifier converts the photocurrent difference into a voltage signal. The resulting voltage signal $\hat{v}(t)$ consists of the detector electronic noise $\hat{v}_{e}(t)$ and the photocurrent-induced voltage $\hat{v}_{r}(t)$:
\begin{equation}
    \hat{v}(t) = \hat{v}_{e}(t) +  \hat{v}_{r}(t),
     \label{eq:v(t)}
\end{equation}
where $\hat{v}_{r}(t)$ is determined by the detector impulse response function $r(t)$ as
\begin{equation}
    \begin{split}
        \hat{v}_{r}(t)&= \int_{-\infty}^{\infty}  dt^{\prime} ~ r( t - t^{\prime}) \hat{i}(t^{\prime}) \\
        &=e \int_{-\infty}^{\infty}  dt^{\prime} ~ r( t - t^{\prime})  |\alpha_{\text{LO}} (t^{\prime})|  \hat{x}_{\theta} (t^{\prime}).
    \end{split} \label{eq:impulse} 
\end{equation}

Next, we describe how the scheme can achieve ultrafast homodyne detection. We consider measurement of the quadrature operator in an ultrafast temporal mode $g(t)$, a real normalized function satisfying $\int_{-\infty}^{\infty} dt g(t)^{2} = 1$. The measurement is repeated with period $T$, which is much longer than the duration of $g(t)$. The ${j}$-th pulse mode is then described by $g(t-{j} T)$, and the associated quadrature operator is given by
\begin{equation}
    \hat{X}_{\theta}^{({j})} = \int_{-\infty}^{\infty} dt^{\prime} g( t^{\prime} - {j} T )~ \hat{x}_{\theta} (t^{\prime}).
\end{equation}
Figure \ref{fig:BHD}(b) illustrates the procedure to measure $\hat{X}_{\theta}^{({j})}$. The LO is prepared as a train of ultrafast pulses with the same temporal mode $g(t)$ and repetition period $T$~\cite{Kumar2012, Cooper2013}: $\alpha_{\text{LO}}(t) = \alpha_{p} \sum_{{j}=-\infty}^{\infty} g(t - {j} T)$. Here, $\alpha_{p} = |\alpha_{p}| e^{i\theta}$ is the pulse amplitude, such that the mean photon number per pulse is $|\alpha_{p}|^{2} = P_{\text{LO}} \lambda T / hc$ ($P_{\text{LO}}$: average LO power, $\lambda$:  wavelength, $h$: the Planck constant, $c$: the speed of light). Upon substituting $\alpha_{\text{LO}}(t)$ into Eq.~(\ref{eq:impulse}), we obtain the photocurrent-induced voltage as
\begin{equation}
	\hat{v}_{r}(t) = e |\alpha_{p}| \sum_{{j}=-\infty} ^{\infty} r(t - {j} T)  \hat{X}_{\theta}^{({j})}.
	\label{eq:vrt}
\end{equation}
In deriving this expression, we used the fact that the detector response function $r(t)$ varies much more slowly than the mode function $g(t)$, so that $r(t - t^{\prime})$ can be approximated by $r(t - {j} T)$ for each ${j}$.
To extract $\hat{X}_{\theta}^{({j})}$ from Eqs.~(\ref{eq:v(t)}) and (\ref{eq:vrt}), we evaluate the weighted integral of $\hat{v}(t)$ over the measurement window $[jT, \, jT+T]$ using a periodic weight function satisfying $w(t) = w(t + j T)$:
\begin{equation}
	\hat{V}^{(j)} = \int_{jT}^{jT+T} dt ~ w(t) \hat{v} (t) = \hat{V}_{e}^{(j)} + \hat{V}_{r}^{(j)},
    \label{eq:weighted_V}
\end{equation}
where
\begin{equation}
    \begin{split}
        \hat{V}_{e}^{(j)} &= \int_{jT}^{jT+T} dt~w(t) \hat{v}_{e} (t) = \int_{0}^{T} dt~w(t) \hat{v}_{e} (t + jT)\\
        \hat{V}_{r}^{(j)} &= \int_{jT}^{jT+T} dt ~ w(t) \hat{v}_{r} (t) = e  |\alpha_{p}| \sum_{{k}=-\infty}^{\infty} \bar{r}_{j,{k}} \hat{X}_{\theta}^{({k})},
	\end{split} \label{eq:Vr} 
\end{equation}
with
\begin{equation}
    \bar{r}_{j,{k}} = \int_{0}^{T} dt~w(t) r(t + (j-{k}) T). \nonumber
\end{equation}
For $j \ne {k}$, $\bar{r}_{j,{k}}$ represents the crosstalk from the ${k}$-th pulse to the $j$-th measurement. When this crosstalk is negligible ($\bar{r}_{j,{k}}=0$), the processed voltage signal $\hat{V}^{(j)}$ reduces to $\hat{V}_{e}^{(j)} + e |\alpha _{p}| \bar{r}_{j,j} \hat{X}_{\theta}^{(j)}$. Importantly, the relative contributions of $\hat{V}_{e}^{(j)}$ and $\hat{V}_{r}^{(j)}$ to $\hat{V}^{(j)}$ depend on the choice of the temporal weight function $w(t)$. In the following section, we present a method for optimizing $w(t)$ to minimize the relative contribution of $\hat{V}_{e}^{(j)}$ for efficient ultrafast homodyne detection.

\section{Optimization of the detection efficiency} \label{section:optimize}

We begin by describing the detection efficiency associated with the SNR of homodyne detection. The SNR is defined as the ratio of the shot-noise variance, obtained from the processed signal $\hat{V}^{(j)}$ for a vacuum input state, to the variance of the electronic noise $\hat{V}_{e}^{(j)}$:
\begin{equation}
	\text{SNR} = \frac{\langle \Delta^2 \hat{V}^{(j)} \rangle} {\langle \Delta^2 \hat{V}_{e}^{(j)} \rangle},
	\label{eq:SNR}
\end{equation}
where $\langle~\rangle$ denotes the expectation value for the vacuum state. The SNR expressed in decibels is commonly referred to as the clearance \cite{Kumar2012}. A finite SNR reduces the effective efficiency of homodyne detection, $\eta_e$, resulting in~\cite{Appel2007}
\begin{equation}
	\quad \eta_{e} = 1 - \frac{1} {\text{SNR}}.
	\label{eq:eff}
\end{equation}
Using $\langle \hat{V}^{(j)} \rangle = \langle \hat{V}_{e}^{(j)} \rangle =0$, the variances in Eq. (\ref{eq:SNR}) can be written as $\langle \Delta^2 \hat{V}^{(j)} \rangle = \langle (\hat{V}^{(j)})^2 \rangle$ and $\langle \Delta^2 \hat{V}_{e}^{(j)} \rangle = \langle (\hat{V}_{e}^{(j)})^2 \rangle$. These quantities are given by
\begin{align}
    \langle (\hat{V}^{(j)})^2 \rangle &= \int_{jT}^{jT+T} \int_{jT}^{jT+T} \! dt\,dt^{\prime} \, w(t) w(t^{\prime}) \, \mathcal{S}(t, t^{\prime}) \label{eq:S} \\
    \langle (\hat{V}_{e}^{(j)})^2 \rangle &= \int_{jT}^{jT+T} \int_{jT}^{jT+T} \! dt\, dt^{\prime} \, w(t) w(t^{\prime}) \, \mathcal{E}(t, t^{\prime}), \label{eq:E}
\end{align}
where $\mathcal{S}(t, t^{\prime}) = \langle \hat{v}(t) \hat{v}(t^{\prime})\rangle$ and $\mathcal{E}(t, t^{\prime}) = \langle \hat{v}_{e}(t) \hat{v}_{e}(t^{\prime})\rangle$, with $\hat{v}(t)$ and $\hat{v}_{e}(t)$ defined in Eq.~(\ref{eq:v(t)}). Similarly,
\begin{equation}
    \langle (\hat{V}_{r}^{(j)})^2 \rangle = \int_{jT}^{jT+T} \int_{jT}^{jT+T} \! dt\,dt^{\prime} \, w(t) w(t^{\prime}) \, \mathcal{R}(t, t^{\prime}), \label{eq:R}
\end{equation}
with $\mathcal{R}(t, t^{\prime}) = \langle \hat{v}_{r}(t) \hat{v}_{r}(t^{\prime})\rangle$. Using $\langle \hat{V}_{r}^{(j)} \hat{V}_{e}^{(j)} \rangle = 0$, we obtain $\mathcal{R}(t, t^{\prime})=\mathcal{S}(t, t^{\prime}) - \mathcal{E}(t, t^{\prime})$. Note that Eqs.~(\ref{eq:S})--(\ref{eq:R}) provide a general formalism without assuming a specific detector model.

For the impulse-response model in Eq. (\ref{eq:vrt}), $\mathcal{R}(t, t^{\prime})$ is given by
\begin{equation}
    \mathcal{R}(t, t^{\prime}) = \eta_{\text{PD}} e^{2} |\alpha_{p}|^{2} \sum_{{k}=-\infty}^{\infty}  r(t - {k} T) r(t^{\prime} - {k} T) \label{eq:Rr},
\end{equation}
where $\eta_{\text{PD}}$ accounts for the quantum efficiency of the photodiodes, and we have used $\langle \hat{X}_{\theta}^{({j})} \hat{X}_{\theta}^{({k})} \rangle = \delta_{j {k}}$, with $\delta_{j {k}}$ denoting the Kronecker delta. In the ideal case where detector response functions from different pulses have negligible overlap,
$\mathcal{R}(t, t^{\prime})$ is given by
\begin{equation}
    \mathcal{R}(t, t^{\prime}) = \eta_{\text{PD}} e^{2} |\alpha_{p}|^{2} r(t - {j} T) r(t^{\prime} - {j} T) \label{eq:Rr2},
\end{equation}
for $t,t^{\prime} \in[jT,jT+T]$. This factorized form corresponds to a rank-one kernel.

Next, we determine the optimal weight function $w(t)$ that maximizes the detection efficiency $\eta_{e}$ in Eq.~(\ref{eq:eff}). To cast this optimization into matrix form, we discretize Eqs. (\ref{eq:S})--(\ref{eq:R}) at $L$ time points $( t_{0}, t_{1}, \dots, t_{L-1} )$, with $t_l=(l/L+j)T$. The voltage signal and the weight function are then represented by the $L$-dimensional vectors $\mathbf{v} = \left(v(t_{0}), v(t_{1}), \dots, v(t_{L-1})\right)^{T}$ and $\mathbf{w} = \left(w(t_{0}), w(t_{1}), \dots, w(t_{L-1})\right)^{T}$, respectively. We also define the shot-noise matrix $\mathbf{S}$, electronic-noise matrix $\mathbf{E}$, and detector response matrix $\mathbf{R}$ by ${S}_{lm} = \mathcal{S} (t_{l}, t_{m})=\langle \hat{v}(t_l) \hat{v}(t_m)\rangle$, $E_{lm} = \mathcal{E} (t_{l}, t_{m})=\langle \hat{v}_e(t_l) \hat{v}_e(t_m)\rangle$, and $R_{lm} = \mathcal{R} (t_{l}, t_{m})=\mathcal{S} (t_{l}, t_{m}) - \mathcal{E} (t_{l}, t_{m})$, respectively. Consequently, the SNR in Eq.~(\ref{eq:SNR}) becomes
\begin{equation}
	\text{SNR}= \frac{\mathbf{w}^{T} \mathbf{S} \mathbf{w}}{\mathbf{w}^{T}\mathbf{E}\mathbf{w}}.
	\label{eq:eff1}
\end{equation}
Finding the optimal weight vector $\mathbf{w}$ that maximizes the SNR is therefore a generalized Rayleigh quotient problem~\cite{Gessner2019, Go2026}. This problem reduces to the generalized eigenvalue equation: $\mathbf{S} \mathbf{w} = \gamma \mathbf{E} \mathbf{w}$, where the largest eigenvalue gives the maximum SNR, and the corresponding eigenvector becomes the optimal weight vector. For example, under the ideal detector response model in Eq.~(\ref{eq:Rr2}) and a white electronic-noise model, the optimal weight vector for the $j$-th measurement is proportional to $\left(r(t_{0}-jT), r(t_{1}-jT), \dots, r(t_{L-1}-jT)\right)^{T}$.

In practice, however, the optimization for Eq.~\eqref{eq:eff1} can yield unreliable results because of sampling errors in the experimental estimation of $\mathbf{S}$ and $\mathbf{E}$. For example, $\mathbf{w}^{T}\mathbf{E}\mathbf{w}$ may become nearly zero while $\mathbf{w}^{T} \mathbf{S} \mathbf{w}$ remains finite, yielding a diverging SNR. To mitigate this issue, we restrict the optimization space of $\mathbf{w}$ to a low-frequency region with a cutoff frequency $f_c$. Specifically, we consider $\mathbf{w}$ expressed as arbitrary superpositions of Fourier-basis vectors $\mathbf{u}^{(n)}$:
\begin{equation}
\mathbf{w}=\sum_{n=0}^{2N} \tilde{w}_{n} \mathbf{u}^{(n)},
\label{eq:wcut}
\end{equation}
where
\begin{equation}
    (\mathbf{u}^{(n)})_l = \begin{cases} 
    \sqrt{{1}/{L}} & (n = 0) \\ 
    \sqrt{{2}/{L}} ~\sin( {(n+1)\pi l}/{L} ) & (n>0, n~\mathrm{odd})  \\ 
    \sqrt{{2}/{L}} ~\cos ( {n\pi  l}/{L} ) & (n>0, n~\mathrm{even}) \end{cases},
\end{equation}
with $2N + 1 < L$. The corresponding cutoff frequency is given by $f_{c} = N/T$. The optimization is therefore reduced to finding the optimal vector $\tilde{\mathbf{w}}= (\tilde{w}_{0}, \tilde{w}_{1}, \dots, \tilde{w}_{2N})^T$, whose dimension is $2N+1$. The optimal weight vector $\mathbf{w}$ is then obtained as $\mathbf{w} = \mathbf{U}_{t} \tilde{\mathbf{w}}$, where
\begin{equation}
    \mathbf{U}_{t} = \left( \mathbf{u}^{(0)}, \mathbf{u}^{(1)}, \mathbf{u}^{(2)}, \dots, \mathbf{u}^{(2N)} \right)
\end{equation}
is the $L \times (2N+1) $ truncated Fourier-basis matrix associated with the low-frequency restriction. The optimization problem in Eq.~\eqref{eq:eff1} then becomes
\begin{equation}
    \max_{\tilde{\mathbf{w}}} \frac{\tilde{\mathbf{w}}^{T} \tilde{\mathbf{S}} \tilde{\mathbf{w}} }{\tilde{\mathbf{w}}^T \tilde{\mathbf{E}} \tilde{\mathbf{w}}},
    \label{eq:eff2}
\end{equation}
where $\tilde{\mathbf{S}}=\mathbf{U}_{t}^T \mathbf{S} \mathbf{U}_{t}$ and $\tilde{\mathbf{E}}=\mathbf{U}_{t}^T \mathbf{E} \mathbf{U}_{t}$ are both $(2N+1) \times (2N+1)$ matrices. The associated generalized eigenvalue equation becomes $\tilde{\mathbf{S}} \tilde{\mathbf{w}} = \gamma \tilde{\mathbf{E}} \tilde{\mathbf{w}}$, which reduces to the standard eigenvalue equation $\tilde{\mathbf{E}}^{-1} \tilde{\mathbf{S}} \tilde{\mathbf{w}} = \gamma \tilde{\mathbf{w}}$ when $\tilde{\mathbf{E}} \succ 0$. The largest eigenvalue gives the maximum SNR, and the corresponding eigenvector $\tilde{\mathbf{w}}$ gives the optimal weight vector $\mathbf{w}$ through $\mathbf{w} = \mathbf{U}_{t} \tilde{\mathbf{w}}$. We apply this method to optimize the SNR for homodyne detection models in Appendix~\ref{appendix:A} and for a real homodyne detector in Section~\ref{section:experiment}.

\begin{figure}[t]
	\centering
	\includegraphics[width=8.5cm]{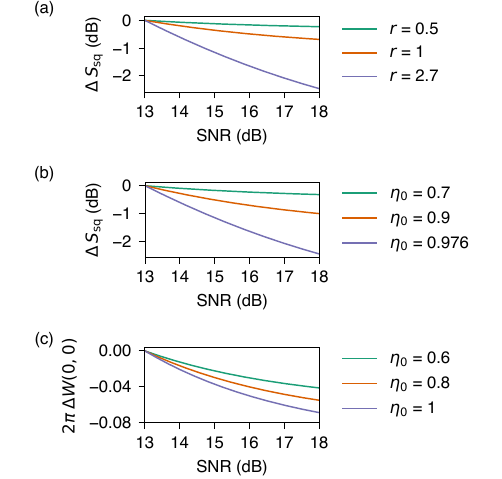}
	\caption{\label{fig:nonclassical} Enhancement of observable nonclassical features with increasing SNR, expressed in dB. The enhancement is defined relative to the value obtained at an initial SNR of 13 dB.
    (a,b) Squeezing enhancement $\Delta S_{\text{sq}}$ in dB for (a) various squeezing parameters $r$ at fixed residual efficiency $\eta_0=0.976$ and (b) various values of $\eta_0$ at fixed $r=2.7$. (c) Enhancement of the Wigner-function negativity at the origin, $\Delta W(0,0)$, for a single-photon state. The quantity is plotted as $2\pi\Delta W(0,0)$; on this scale, a pure single-photon state has $2\pi W(0,0) = -1$.}
\end{figure}

We further explore how the SNR improvement can facilitate the observation of nonclassical features in quantum states. As shown in Eq.~\eqref{eq:eff}, the SNR improvement increases the detection efficiency $\eta_{e}$, thereby enhancing the overall optical efficiency, $\eta=\eta_{e}\eta_{0}$. Here, $\eta_{0}$ denotes the residual efficiency, which accounts for all other efficiency factors, such as the photodiode efficiency, optical propagation efficiency, and mode-matching efficiency.

For example, the SNR improvement enhances the observable noise reduction of a squeezed vacuum. The measured noise variance is given by $S_\text{sq}=(1 - \eta_{e} \eta_{0}) + \eta_{e} \eta_{0} \exp (-2r)$, where $r > 0$ is the squeezing parameter, and a higher $\eta_{e}$ enables clearer observation of the intrinsic noise reduction, $\exp (-2r)$. Figure~\ref{fig:nonclassical}(a) and (b) show the squeezing enhancement, $\Delta S_{\text{sq}}$, as the SNR increases. For an SNR increase from 13~dB to 18~dB, with $\eta_0$ fixed at 0.976, the squeezing enhancement is 0.23~dB, 0.68~dB, and 2.46~dB for $r=0.5$, $r=1$, and $r=2.7$, respectively, as shown in Fig.~\ref{fig:nonclassical}(a). The most stringent condition considered here, $r=2.7$ and $\eta_0 = 0.976$, is also experimentally feasible, as demonstrated in Ref.~\cite{Vahlbruch2016}. Similarly, at fixed $r=2.7$, the increased SNR enhances the observable squeezing by $0.32$ dB, $1.00$ dB, and $2.46$ dB for $\eta_0 = 0.7$, $\eta_0 = 0.9$, and $\eta_0 = 0.976$, respectively, as shown in Fig.~\ref{fig:nonclassical}(b). In addition, the SNR improvement enables clearer observation of Wigner-function negativity of non-Gaussian states~\cite{Ra2020, Endo2023}. As an example, Fig.~\ref{fig:nonclassical}(c) shows the enhancement of the observable Wigner-function negativity at the origin, $\Delta W(0,0)$, for a single-photon state. 

\section{Experiment}\label{section:experiment}

\begin{figure}[t]
	\centering
	\includegraphics[width=8.5cm]{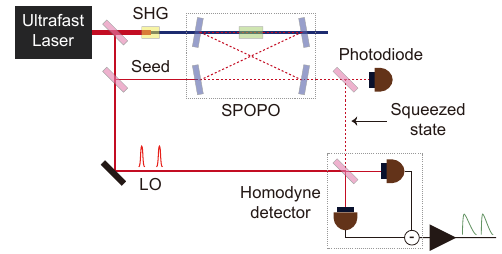}
	\caption{Experimental setup.  A Ti:sapphire laser (75~fs pulse width, 80~MHz repetition rate, 800~nm central wavelength) is employed as an ultrafast light source. A portion of the light is converted to 400~nm via SHG and used as the pump for the SPOPO to generate a pulsed squeezed state. A weak coherent seed beam is additionally injected into the SPOPO to lock the relative phase between the squeezed state and the LO. The homodyne detector measures the quadratures of the vacuum or squeezed vacuum.
    }
	\label{fig:setup}
\end{figure}

We experimentally demonstrate the optimization of homodyne detection efficiency and the resulting enhancement of squeezing measurements. Figure~\ref{fig:setup} shows the experimental setup for generating ultrafast squeezed vacuum and measuring it with an ultrafast homodyne detector. We use a Ti:sapphire femtosecond laser (Chameleon Vision-S, Coherent) with a central wavelength of 800~nm, a pulse duration of 75~fs, and a repetition rate of 80~MHz. A frequency-doubled beam is prepared via second-harmonic generation (SHG) in a 2-mm-thick $\text{LBO}$ crystal. The output 400~nm pulses serve as the pump for a synchronously pumped optical parametric oscillator (SPOPO) containing a 2-mm-thick $\text{BiBO}$ crystal, which generates ultrafast squeezed vacuum. The relative phase between the squeezed vacuum and the pump is locked by injecting a weak seed beam into the SPOPO and monitoring the tapped output power at a photodiode. The squeezed vacuum then interferes with the LO, whose spectral full width at half maximum is 3.3~nm, at a 50:50 beam splitter. The output beams are detected by two photodiodes (Hamamatsu S5971, quantum efficiency: 95.4\% at 800~nm) in the homodyne detector. The relative phase between the squeezed vacuum and the LO is locked using the near-DC interference signal between the seed beam from the SPOPO and the LO.

The homodyne output signal is recorded with an oscilloscope (WaveRunner 8108, Teledyne LeCroy). During acquisition, the signal is continuously sampled at 10~GSa/s, yielding 125 samples for each measurement period, $T=12.5$~ns. Because the sampling rate is not an exact integer multiple of the pulse repetition rate, the measurement window drifts slightly relative to the pulse train, producing effective timing jitter. We mitigate this effect using a polyphase finite-impulse-response resampling algorithm: the 125 samples in each period are interpolated to 1,000 samples, the small temporal drift is compensated, and the data are then downsampled back to 125 samples. This process synchronizes each optical pulse with the measurement window while preserving the detector bandwidth and suppressing aliasing. Finally, a voltage offset at each sampling point is subtracted, yielding the voltage signal $v(t)$ for subsequent analysis.

\begin{figure}[t]
	\centering
	\includegraphics[width=8.5cm]{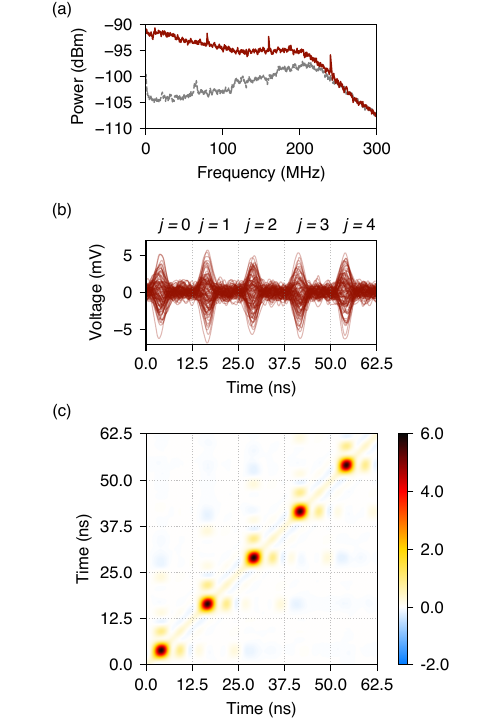}
	\caption{Homodyne detector characteristics. (a) Power spectrum of shot noise for 4~mW LO power (red) and of electronic noise (gray) measured with a spectrum analyzer in the range of 10~kHz to 300~MHz. (b) Example of 500 shot-noise pulses rearranged into five-period blocks and overlaid. (c) Shot-noise matrix $\mathbf{S}$ for 4~mW LO power over five periods.}
	\label{fig:HDcharacteristics}
\end{figure}

We first characterize the homodyne detector used in the experiment. In the spectral domain (Fig.~\ref{fig:HDcharacteristics}(a)), the shot-noise measurement shows that the homodyne detector has a broad spectral response reaching nearly 200~MHz, well above the pulse repetition rate of 80 MHz. The electronic noise remains lower than the shot noise, with a maximum clearance of about 13~dB. 
We also investigate the temporal characteristic using five consecutive pulses, from $j=0$ to $j=4$, as shown in Fig.~\ref{fig:HDcharacteristics}(b). Each pulse response is well confined within the corresponding measurement window. Figure~\ref{fig:HDcharacteristics}(c) shows the associated shot-noise matrix $\mathbf{S}$. A strong correlation appears as a localized peak in each period, dominating the other correlations. This strong correlation can be exploited to optimize the homodyne detection efficiency.

\begin{figure*}[p]
	\centering
	\includegraphics[width=17.8cm]{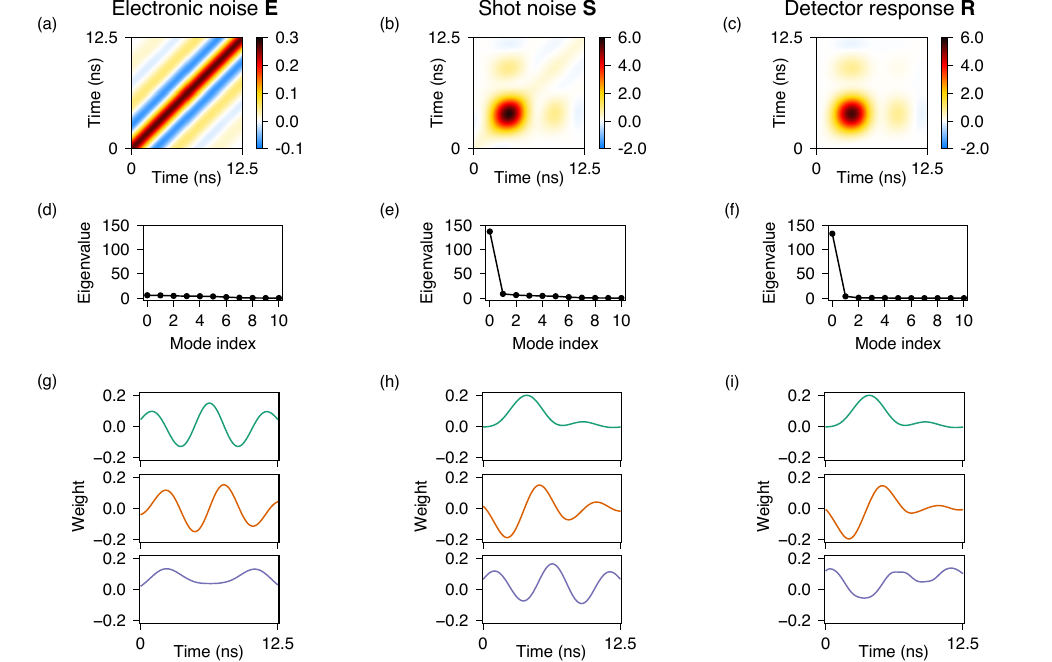}
	\caption{Analysis of homodyne detector signals. (a)--(c) Autocorrelation matrices: (a) electronic-noise matrix $\mathbf{E}$, (b) shot-noise matrix $\mathbf{S}$, and (c) detector response matrix $\mathbf{R}$. Matrices in (b) and (c) are obtained at an LO power of 4 mW, and all matrices are determined from 80,000 repeated measurements. (d)--(f) Eigenvalues of (d) $\mathbf{E}$, (e) $\mathbf{S}$, and (f) $\mathbf{R}$, shown in descending order. (g)--(i) The three leading eigenvectors of (g) $\mathbf{E}$, (h) $\mathbf{S}$, and (i) $\mathbf{R}$.}
	\label{fig:cov}
    \vspace{1cm}
	\includegraphics[width=17.8cm]{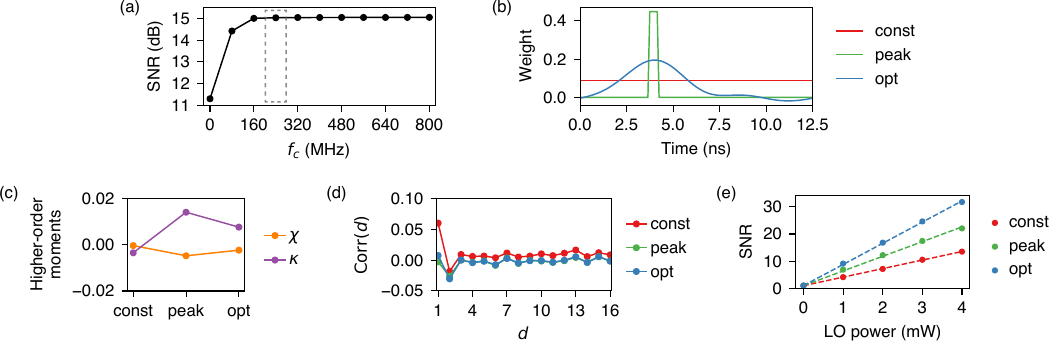}
	\caption{
    Performance comparison of various weight vectors.
    (a) SNR as a function of the cutoff frequency $f_{c}$. The LO power is 4 mW. The gray dashed box indicates $f_{c}$ = 240~MHz, which is used to determine the optimal weight vector shown in (b). (b) Temporal profiles of the three weight vectors: constant (const, red), peak-shaped (peak, green; 0.5~ns peak), and optimal (opt, blue). Their performance is compared in (c)--(e). (c) Skewness ($\chi$) and excess kurtosis ($\kappa$) of the shot-noise measurement. (d) Correlation coefficient, $\text{Corr}(d)$, between pulses separated by $d$ periods, as defined in Eq.~(\ref{eq:corr}). (e) SNR as a function of the LO power. The optimal weight vectors for LO powers of 1--3 mW are determined in the same manner as in the 4-mW case. The dots show the experimental data, and the dashed lines represent linear fits to $\text{SNR} = a P_{\text{LO}} + 1$, where $P_{\text{LO}}$ is the LO power and $a$ is the linear coefficient. The $R^{2}$ values of the fits are 0.999698 for const, 0.997267 for peak, and 0.999355 for opt.
    }
	\label{fig:weight}
    
\end{figure*}

Next, we analyze the detector signal $v(t)$ within a single period using autocorrelation matrices. Figures~\ref{fig:cov}(a)--(c) show the corresponding electronic-noise matrix $\mathbf{E}$, shot-noise matrix $\mathbf{S}$, and the detector response matrix $\mathbf{R}$, respectively. The contribution of $\mathbf{R}$ to $\mathbf{S}$ dominates over that of $\mathbf{E}$, enabling a high SNR in homodyne detection. In terms of their temporal structures, $\mathbf{E}$ depends only on the time difference $t_l -t_m$, reflecting the stationarity of the electronic noise. By contrast, $\mathbf{S}$ and $\mathbf{R}$ exhibit localized peaks in time, with correlations extending over nearby off-diagonal elements. This suggests that the matrices are dominated by a few leading eigenvectors, which becomes clearer via the eigenvalue decomposition. In Fig.~\ref{fig:cov}(d)--(f), the eigenvalues of $\mathbf{E}$ decrease gradually, whereas those of $\mathbf{S}$ and $\mathbf{R}$ are dominated by single large values. In particular, the dominance of the leading eigenvalue of $\mathbf{R}$ is consistent with the ideal detector response model in Eq.~\eqref{eq:Rr2}. Figures~\ref{fig:cov}(g)--(i) show the three leading eigenvectors (1st: green, 2nd: orange, 3rd: blue) of $\mathbf{E}$, $\mathbf{S}$ and $\mathbf{R}$, respectively. The leading eigenvector of $\mathbf{R}$ reveals the dominant response function $r(t)$ of the homodyne detector.

We now determine the optimal weight vector $\mathbf{w}$ from the obtained autocorrelation matrices, following the method developed in Section~\ref{section:optimize}. Figure~\ref{fig:weight}(a) shows the maximum SNR as a function of the cutoff frequency $f_{c}$ in Eq.~(\ref{eq:wcut}). The case $f_{c}=0$ corresponds to the constant weight vector shown in Fig.~\ref{fig:weight}(b), yielding an SNR of 11.3~dB. This constant-weight strategy is equivalent to the conventional area-integration method~\cite{Zavatta2006, Okubo2008, Kumar2012, Cooper2013, Kouadou2026}. Increasing the cutoff frequency substantially improves the SNR, reaching 15.0~dB at $f_{c}=3/T$ (240 MHz). The corresponding optimal weight vector is shown in Fig.~\ref{fig:weight}(b), having a similar shape to the first eigenvector in Fig.~\ref{fig:cov}(i). A further increase in $f_{c}$ does not lead to a noticeable improvement because of the finite spectral range of the homodyne detector (Fig.~\ref{fig:HDcharacteristics}(a)). For comparison, we also examine another common strategy that uses the voltage at the peak of the detector response~\cite{Hansen2001, Du2018}. Figure~\ref{fig:weight}(b) shows the corresponding peak-shaped weight vector, which yields an SNR of 13.4~dB.

We further compare the performance of the three weight vectors: constant, peak-shaped, and optimal. First, we evaluate the skewness ($\chi$) and excess kurtosis $(\kappa)$ of the shot-noise data obtained with each weight vector. As shown in Fig.~\ref{fig:weight}(c), both quantities are close to zero for all three cases, consistent with the expected Gaussian statistics, although the largest deviation is observed for the peak-shaped weight vector. Next, we examine the correlations between $\hat{V}^{(j)}$ and $\hat{V}^{(j+d)}$ for pulses separated by $d$ periods. The corresponding correlation coefficient, $\text{Corr}(d)$, is defined as
\begin{equation}
	\text{Corr}(d) = \frac{\langle \Delta\hat{V}^{(j)} \Delta\hat{V}^{(j+d)} \rangle} {\sqrt{\langle \Delta^2\hat{V}^{(j)}\rangle \langle \Delta^2\hat{V}^{(j+d)} \rangle}},
    \label{eq:corr}
\end{equation}
where the expectation values are obtained with the vacuum input state. Figure~\ref{fig:weight}(d) shows the correlation coefficients for the three weight vectors. The optimal and peak-shaped weight vectors exhibit negligible correlations, as desired, whereas the constant weight vector shows a slight correlation of 6.0\% at $d=1$. We further verify the linearity of the shot noise variance as the LO power is increased. For all three cases, a clear linear trend is observed in Fig.~\ref{fig:weight}(e), and the highest SNR is obtained with the optimal weight vector.

\begin{figure*}[t]
	\centering
	\includegraphics[width=17.8cm]{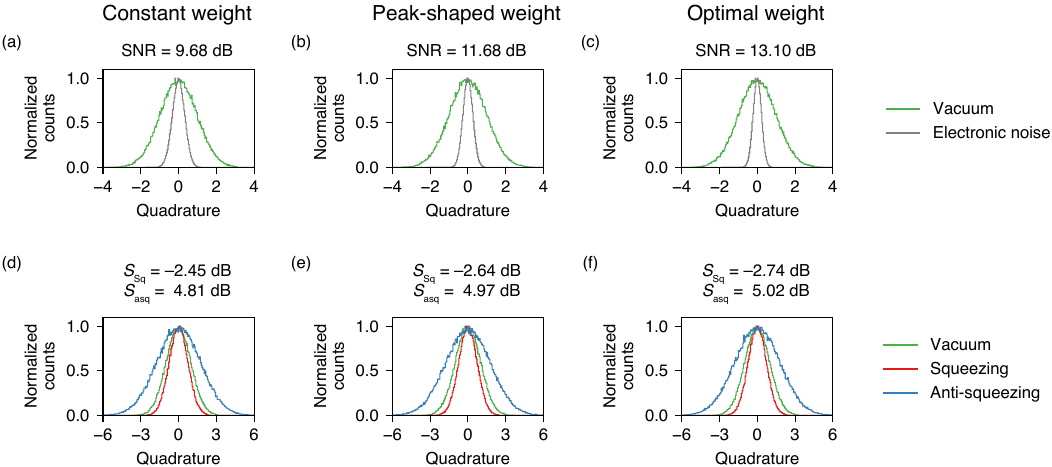}
	\caption{Histograms of quadrature outcomes obtained with various weight vectors.
    (a)--(c) Histograms of the vacuum measurement (green) and electronic noise (gray). (d)--(f) Histograms of the squeezing measurement (red) and the anti-squeezing measurement (blue), compared with the vacuum measurement (green). The weight vectors used to obtain the quadrature outcomes are the constant weight for (a,d), the peak-shaped weight for (b,e), and the optimal weight for (c,f).    
    The homodyne-detection SNRs for (a)--(c) and the squeezing level $S_{\text{sq}}$ and anti-squeezing level $S_{\text{asq}}$ for (d)--(f) are indicated at the top of each panel. Each histogram is obtained from 125,000 quadrature outcomes at a 1.5~MHz sideband frequency. All histograms are normalized to have a maximum value of unity. The uncertainties of the squeezing and anti-squeezing levels are $\pm 0.048$~dB, estimated from the $F$-distribution at a 95\% confidence level.
    }
	\label{fig:squeezing}
\end{figure*}

Finally, we demonstrate the enhancement of squeezing measurements achieved with the optimal weight vector.
We measure the squeezed light generated by the SPOPO at a sideband frequency of $f_s$ = 1.5~MHz by applying a sinusoidal weight to the pulse-by-pulse signals $V^{(j)}$:
\begin{equation}
\Phi=\sum_{j=0}^{J-1} \sin (2 \pi f_s jT) V^{(j)}.
\end{equation}
Here, $V^{(j)}=\sum_{l=0}^{L-1} w_{l} v(t_{l})$ is the processed signal for the $j$-th period, obtained by applying the weights $w_{l}=(\mathbf{w})_{l}$ to the voltage signals $v(t_{l})$, where $t_{l}=(l / L+j)T$. In the experiment, we use an LO power of 2.5 mW and determine the optimal weight vector using the same procedure described above. We set $J = 640$, corresponding to 12 sinusoidal periods at $f_s$=1.5 MHz, and $L=125$. Each value of $\Phi$ is normalized such that the vacuum variance is unity, yielding a quadrature outcome at the sideband frequency.

Figure~\ref{fig:squeezing} shows the experimental results. We record the voltage signal $v(t_l)$ separately for the vacuum and squeezed-vacuum states and, for each state, process the same signal using three different weight vectors to compare their performance. The vacuum measurement results are shown in Fig.~\ref{fig:squeezing}(a--c). Among the three weight vectors, the optimal one gives the highest SNR, as reflected in the narrowest electronic-noise distribution. Figure~\ref{fig:squeezing}(d--f) shows the squeezed-vacuum measurement results obtained at the squeezing and anti-squeezing phases. Both the squeezing and anti-squeezing levels improve sequentially for the constant, peak-shaped, and optimal weight vectors.

We further confirm that the observed improvements in squeezing and anti-squeezing measurements are consistent with theoretical predictions. From Fig.~\ref{fig:squeezing}(d), the squeezing and anti-squeezing levels, $S_{\text{sq}}$ and $S_{\text{asq}}$, allow us to determine the overall efficiency $\eta$ and the squeezing parameter $r>0$ using 
\begin{align}
    S_{\text{sq}} &= (1 - \eta) + \eta \exp(-2r) \nonumber \\
    S_{\text{asq}} &= (1 - \eta) + \eta \exp(2r). \label{eq:sqz}
\end{align}
The improvement factor of the detection efficiency due to the increased SNR is evaluated from Eq.~(\ref{eq:eff}), which is denoted by $\zeta$. The corresponding improved squeezing and anti-squeezing levels are then predicted by replacing $\eta$ with $\zeta\eta$ in Eq.~(\ref{eq:sqz}). The predicted squeezing and anti-squeezing levels are $-$2.60 dB and 4.93 dB in Fig.~\ref{fig:squeezing}(e), and $-$2.67 dB and 4.99 dB in Fig.~\ref{fig:squeezing}(f), agreeing well with the experimental results.

\section{Conclusion}

We have demonstrated a method for enhancing the SNR of ultrafast homodyne detection by optimizing the temporal weight used to extract quadrature outcomes from detector signals. The optimal weight is determined by analyzing the autocorrelation matrices of the shot noise and the electronic noise and solving the associated generalized Rayleigh quotient problem. Experimentally, the optimal weight increases the SNR by 135\% relative to the constant weight and by 44\% relative to the peak-shaped weight, leading to improved detection efficiency. It also reduces pulse-to-pulse correlations more effectively than the constant weight, while yielding shot-noise statistics closer to a Gaussian distribution than the peak-shaped weight. Finally, we show that this improvement in homodyne detection enhances the observed squeezing and anti-squeezing levels. Our approach to SNR enhancement provides a practical route to improving quantum measurements in which readout noise plays a central role~\cite{Gambetta2007, Eaton2023}.

\section*{Acknowledgments}
This work was supported by the Ministry of Science and ICT (MSIT) of Korea (RS-2025-00562372, RS-2024-00408271, RS-2024-00442762, RS-2023-NR119925) under the Information Technology Research Center (ITRC) support program (IITP-2026-RS-2020-II201606) and Institute of Information \& Communications Technology Planning \& Evaluation (IITP) grant (RS-2025-25464959, RS-2022-II221029).

\appendix
\renewcommand{\thefigure}{S\arabic{figure}}
\setcounter{figure}{0}

\section{Simulation of weight optimization} \label{appendix:A}
\begin{figure*}[p]
	\centering
	\includegraphics[width=17.8cm]{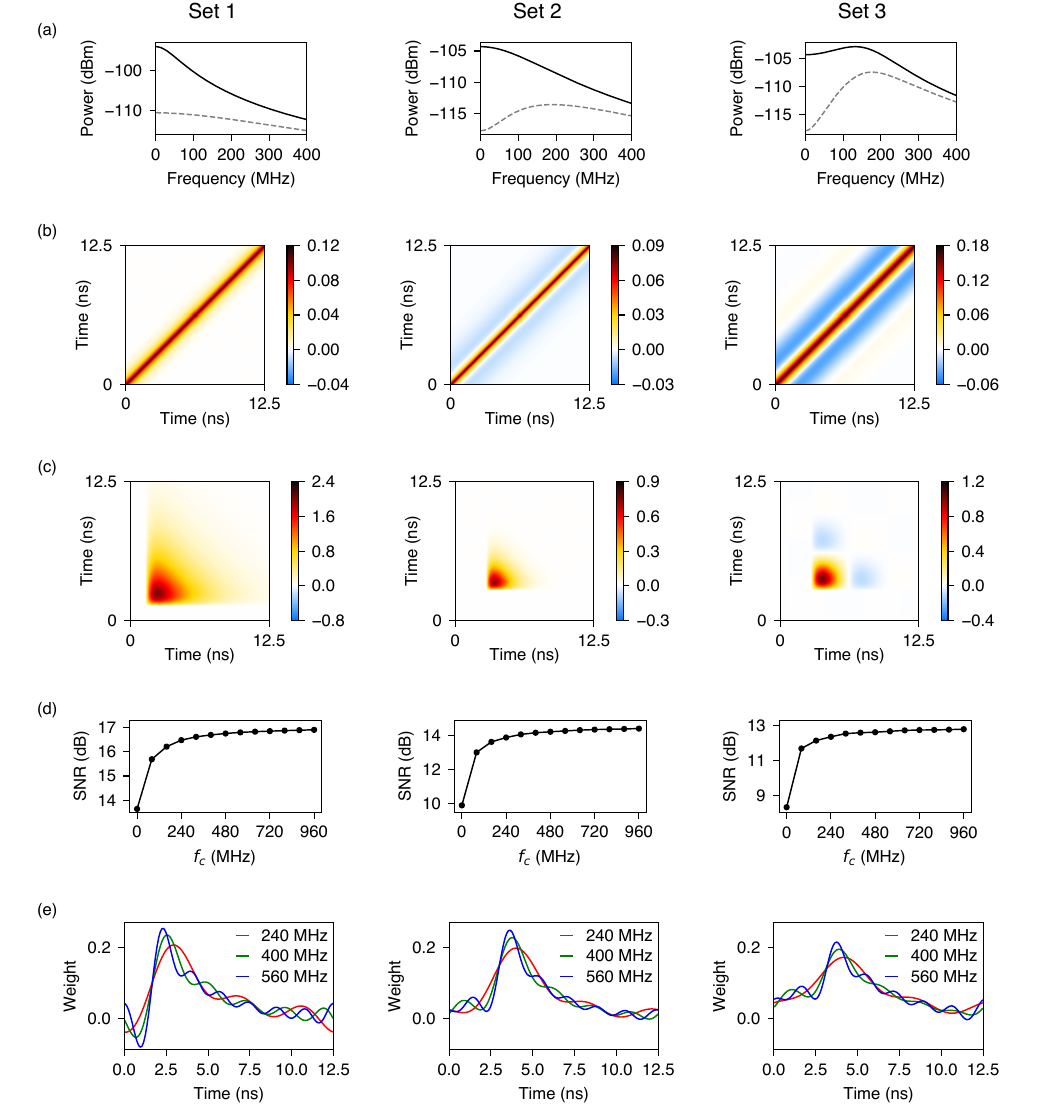}
	\caption{\label{fig:model} Simulation results of weight-vector optimization for three different parameter sets of a homodyne detector (left: Set 1, center: Set 2, right: Set 3, as defined in Table~\ref{tab:model}). (a) Power spectrum of the shot noise for LO power of 2.7~mW with $\eta_{\text{PD}} = 0.9$ (black), and the electronic noise (gray, dashed). Both are measured with 10~kHz resolution bandwidth. (b) Electronic-noise matrices $\mathbf{E}$. (c) Detector response matrices $\mathbf{R}$ at $\eta_{\text{PD}} = 0.9$ and LO power of 2.7~mW. (d) SNR as a function of the cutoff frequency up to $f_{c}$ = 960~MHz. (e) Optimal weight vectors for $f_{c}$ = 240~MHz (red), 400~MHz (green), and 560~MHz (blue).}
\end{figure*}

We simulate the SNR improvement achieved by the optimization method described in Section~\ref{section:optimize}, using a realistic homodyne circuit model. First, we assume that the electronic noise is stationary, meaning that $E_{lm} = \left\langle \hat{v}_{e}(t_{l})\hat{v}_{e}(t_{m}) \right\rangle = \left\langle \hat{v}_{e}(t)\hat{v}_{e}(t + \Delta t_{lm}) \right\rangle$, where $\Delta t_{lm}=t_{m}-t_{l}$. Thus, $E_{lm}$ can be regarded as an autocorrelation matrix of the electronic noise that depends only on $\Delta t_{lm}$. According to the Wiener--Khinchin theorem, this stationary autocorrelation matrix is the inverse Fourier transform of the power spectral density of the electronic noise $S_{e}(f)$:
\begin{equation}
     E_{lm} = \mathcal{F}^{-1} \left[S_{e}\right](\Delta t_{lm}),
     \label{eq:freq1}
\end{equation}
where $\mathcal{F}^{-1}$ denotes the inverse Fourier transform, defined as $\mathcal{F}^{-1}[G](t) = \int_{-\infty}^{\infty} df G(f) e^{i2\pi ft}$. Furthermore, $R_{lm}$ is proportional to the product of impulse response functions at different times, provided that detector response functions from different pulses have negligible overlap (see Eq. (\ref{eq:Rr2})). The impulse response in the frequency domain is given by the transfer function $\tilde{r}(f) = \mathcal{F}[r](f)$. Combining this expression with Eq.~\eqref{eq:Rr2}, $R_{lm}$ is rewritten as
\begin{equation}
     R_{lm} = \eta_{\text{PD}} e^{2} |\alpha_{p}|^{2} \ \mathcal{F}^{-1}\!\left[ \tilde{r} \right] (t_{l} -jT) \ \mathcal{F}^{-1}\!\left[ \tilde{r} \right] (t_{m} -jT),
     \label{eq:freq2}
\end{equation}
for $t_{l}, t_{m} \in [jT,jT+T]$.

We use the homodyne circuit model in Ref.~\cite{Masalov2017}. In this model, the photodiode is regarded as a current source with an additional variable capacitor $C_{p}$ in parallel. The power spectral density of the electronic noise $S_{e}(f)$ and the response function $\tilde{r}(f)$ of this circuit are given by
\begin{align}
	S_{e}(f) &= \dfrac{n_{c} + n_{f} \bar{f}^{2}}{1+(p^{2}-2) \bar{f}^{2} + \bar{f}^{4}} \label{eq:model_elec} \\
	\tilde{r}(f) &= \dfrac{R_{f}}{1+ip \bar{f} - \bar{f}^{2}} 
	\label{eq:model_resp},
\end{align}
respectively, where a normalized frequency $\bar{f}$ is defined as $\bar{f} = f / f_{0}$, $p=f_{0}\left(2\pi R_{f}C_{f} +1/\text{GBW} \right)$ and $f_{0}=\sqrt{\text{GBW} / (2\pi R_{f} C_{t})}$. The noise coefficients are given by $n_{c} = e_{n}^{2} + 4k_{B} T_{e} R_{f} +i_{n}^{2} R_{f}^{2}$, and $n_{f} = 4\pi^{2} e_{n}^{2} R_{f}^{2} C_{t}^{2} f_{0}^{2}$, where $k_{B}$ is the Boltzmann constant and $T_{e}$ is the temperature. Here, GBW is the gain-bandwidth product, and $i_{n}$ and $e_{n}$ are the input noise current density and voltage density of the op-amp, respectively. The circuit parameters $R_{f}$ and $C_{f}$ represent the feedback resistance and the capacitance of the transimpedance amplifier, respectively. $C_{a}$ is the input capacitance of the op-amp, and $C_{t} = 2C_{p} + C_{f} + C_{a}$ is the total capacitance.

We simulate this homodyne detector model using an OPA847 op-amp and varying $R_{f}$, $C_{f}$, and $C_{p}$, which are given in Table~\ref{tab:model}. The differences in the characteristic frequency $f_{0}$ and damping parameter $p$ determine the impulse responses of each parameter set. We set $\eta_{\text{PD}} = 0.9$, the LO power to 2.7~mW, and $T_{e} = 298.15~\text{K}~(25~^{\circ} \text{C})$. The following values are taken from the datasheet of the OPA847: $i_{n} = 2.5~\text{pA}/ \sqrt{\text{Hz}}$, $e_{n} = 0.89~\text{nV}/ \sqrt{\text{Hz}}$, $\text{GBW} = 3.9~\text{GHz}$, and $C_{a} = 2~\text{pF}$. Using these parameters and Eqs.~\eqref{eq:model_elec}--\eqref{eq:model_resp}, we first plot the spectra of shot noise and electronic noise, shown in Fig.~\ref{fig:model}(a). Their shapes depend on $f_{0}$ and $p$: $f_{0}$ determines the time and frequency scale, whereas $p$ sets the damping behavior.

\begingroup
\renewcommand{\arraystretch}{1.1}
\begin{table}[h]
    \caption{\label{tab:model} Circuit parameter sets used in homodyne detector simulations.}
    \begin{ruledtabular}
        \begin{tabular}{cccc}
        Parameter & Set 1 & Set 2 & Set 3 \\ \hline
        $R_{f}$ (k$\Omega$) & 3.3 & 1.0 & 1.0 \\
        $C_{f}$ (pF)        & 1.0 & 1.5 & 1.0 \\
        $C_{p}$ (pF)        & 4.0 & 5.0 & 8.0 \\
        $f_{0}$ (MHz)       & 130.8 & 214.4 & 180.7 \\
        $p$                 & 2.74 & 2.08 & 1.18 \\
        \end{tabular}
    \end{ruledtabular}
\end{table}
\endgroup
\noindent
For larger $p$, $S_{e}(f)$ and $|\tilde{r} (f)|^{2}$ monotonically decrease with frequency, whereas they show non-zero maximum peaks as $p$ decreases. Moreover, in the time domain, both the autocorrelation of the electronic noise and the detector response are underdamped for $p < 2$, critically damped for $p = 2$, and overdamped for $p > 2$. These behaviors are reflected in the structures of $\mathbf{E}$ and $\mathbf{R}$, shown in Figs.~\ref{fig:model}(b) and (c), respectively. Since $E_{lm}$ is determined by the autocorrelation function, whereas $R_{lm}$ is proportional to the product of the response functions at different times, $R_{lm}$ is more localized in time.

Using $\mathbf{E}$ and $\mathbf{S} = \mathbf{E} + \mathbf{R}$, we evaluate the SNR as a function of the cutoff frequency $f_{c}$, shown in Fig.~\ref{fig:model}(d). The SNRs of the three parameter sets obtained with the constant weight vector are 13.8~dB, 9.9~dB, and 8.3~dB, respectively. These values increase to 16.6~dB, 13.9~dB, and 12.4~dB for $f_{c}$ = 240~MHz, and to 16.9~dB, 14.3~dB, and 12.7~dB for $f_{c}$ = 560~MHz. Figure~\ref{fig:model}(e) shows the corresponding optimal weight vectors for three cutoff frequencies: $f_{c}$ = 240~MHz, $f_{c}$ = 400~MHz, and $f_{c}$ = 560~MHz for each circuit parameter set. As $f_{c}$ increases, high-frequency oscillations are observed in the optimal weight vectors.

\bibliographystyle{naturemag}
\bibliography{ref}

\begin{thebibliography}{10}
\expandafter\ifx\csname url\endcsname\relax
  \def\url#1{\texttt{#1}}\fi
\expandafter\ifx\csname urlprefix\endcsname\relax\def\urlprefix{URL }\fi
\providecommand{\bibinfo}[2]{#2}
\providecommand{\eprint}[2][]{\url{#2}}

\bibitem{Larsen2019}
\bibinfo{author}{Larsen, M.~V.}, \bibinfo{author}{Guo, X.},
  \bibinfo{author}{Breum, C.~R.}, \bibinfo{author}{Neergaard-Nielsen, J.~S.} \&
  \bibinfo{author}{Andersen, U.~L.}
\newblock \bibinfo{title}{{Deterministic generation of a two-dimensional
  cluster state}}.
\newblock \emph{\bibinfo{journal}{Science}} \textbf{\bibinfo{volume}{366}},
  \bibinfo{pages}{369--372} (\bibinfo{year}{2019}).

\bibitem{Asavanant2019}
\bibinfo{author}{Asavanant, W.} \emph{et~al.}
\newblock \bibinfo{title}{{Generation of time-domain-multiplexed
  two-dimensional cluster state}}.
\newblock \emph{\bibinfo{journal}{Science}} \textbf{\bibinfo{volume}{366}},
  \bibinfo{pages}{373--376} (\bibinfo{year}{2019}).

\bibitem{Guo2020}
\bibinfo{author}{Guo, X.} \emph{et~al.}
\newblock \bibinfo{title}{{Distributed quantum sensing in a continuous-variable
  entangled network}}.
\newblock \emph{\bibinfo{journal}{Nat. Phys.}} \textbf{\bibinfo{volume}{16}},
  \bibinfo{pages}{281--284} (\bibinfo{year}{2020}).

\bibitem{Kovalenko2021}
\bibinfo{author}{Kovalenko, O.} \emph{et~al.}
\newblock \bibinfo{title}{{Frequency-multiplexed entanglement for
  continuous-variable quantum key distribution}}.
\newblock \emph{\bibinfo{journal}{Photonics Res.}}
  \textbf{\bibinfo{volume}{9}}, \bibinfo{pages}{2351--2359}
  (\bibinfo{year}{2021}).

\bibitem{Deng2023}
\bibinfo{author}{Deng, Y.-H.} \emph{et~al.}
\newblock \bibinfo{title}{{Gaussian Boson Sampling with
  Pseudo-Photon-Number-Resolving Detectors and Quantum Computational
  Advantage}}.
\newblock \emph{\bibinfo{journal}{Phys. Rev. Lett.}}
  \textbf{\bibinfo{volume}{131}}, \bibinfo{pages}{150601}
  (\bibinfo{year}{2023}).

\bibitem{AghaeeRad2025}
\bibinfo{author}{Aghaee~Rad, H.} \emph{et~al.}
\newblock \bibinfo{title}{{Scaling and networking a modular photonic quantum
  computer}}.
\newblock \emph{\bibinfo{journal}{Nature}} \textbf{\bibinfo{volume}{638}},
  \bibinfo{pages}{912--919} (\bibinfo{year}{2025}).

\bibitem{Jia2025}
\bibinfo{author}{Jia, X.} \emph{et~al.}
\newblock \bibinfo{title}{{Continuous-variable multipartite entanglement in an
  integrated microcomb}}.
\newblock \emph{\bibinfo{journal}{Nature}} \textbf{\bibinfo{volume}{639}},
  \bibinfo{pages}{329--336} (\bibinfo{year}{2025}).

\bibitem{Liu2025}
\bibinfo{author}{Liu, Z.-H.} \emph{et~al.}
\newblock \bibinfo{title}{{Quantum learning advantage on a scalable photonic
  platform}}.
\newblock \emph{\bibinfo{journal}{Science}} \textbf{\bibinfo{volume}{389}},
  \bibinfo{pages}{1332--1335} (\bibinfo{year}{2025}).

\bibitem{Andersen2016}
\bibinfo{author}{Andersen, U.~L.}, \bibinfo{author}{Gehring, T.},
  \bibinfo{author}{Marquardt, C.} \& \bibinfo{author}{Leuchs, G.}
\newblock \bibinfo{title}{{30 years of squeezed light generation}}.
\newblock \emph{\bibinfo{journal}{Phys. Scr.}} \textbf{\bibinfo{volume}{91}},
  \bibinfo{pages}{053001} (\bibinfo{year}{2016}).

\bibitem{Casacio2021}
\bibinfo{author}{Casacio, C.~A.} \emph{et~al.}
\newblock \bibinfo{title}{{Quantum-enhanced nonlinear microscopy}}.
\newblock \emph{\bibinfo{journal}{Nature}} \textbf{\bibinfo{volume}{594}},
  \bibinfo{pages}{201--206} (\bibinfo{year}{2021}).

\bibitem{Nehra2022}
\bibinfo{author}{Nehra, R.} \emph{et~al.}
\newblock \bibinfo{title}{{Few-cycle vacuum squeezing in nanophotonics}}.
\newblock \emph{\bibinfo{journal}{Science}} \textbf{\bibinfo{volume}{377}},
  \bibinfo{pages}{1333--1337} (\bibinfo{year}{2022}).

\bibitem{Rasputnyi2024}
\bibinfo{author}{Rasputnyi, A.} \emph{et~al.}
\newblock \bibinfo{title}{{High-harmonic generation by a bright squeezed
  vacuum}}.
\newblock \emph{\bibinfo{journal}{Nat. Phys.}} \textbf{\bibinfo{volume}{20}},
  \bibinfo{pages}{1960--1965} (\bibinfo{year}{2024}).

\bibitem{Herman2025}
\bibinfo{author}{Herman, D.~I.} \emph{et~al.}
\newblock \bibinfo{title}{{Squeezed dual-comb spectroscopy}}.
\newblock \emph{\bibinfo{journal}{Science}} \textbf{\bibinfo{volume}{387}},
  \bibinfo{pages}{653--658} (\bibinfo{year}{2025}).

\bibitem{Sennary2025}
\bibinfo{author}{Sennary, M.} \emph{et~al.}
\newblock \bibinfo{title}{{Attosecond quantum uncertainty dynamics and
  ultrafast squeezed light for quantum communication}}.
\newblock \emph{\bibinfo{journal}{Light Sci. Appl.}}
  \textbf{\bibinfo{volume}{14}}, \bibinfo{pages}{350} (\bibinfo{year}{2025}).

\bibitem{Cimini2026}
\bibinfo{author}{Cimini, V.} \emph{et~al.}
\newblock \bibinfo{title}{{Large-scale quantum reservoir computing using a
  Gaussian Boson Sampler}}.
\newblock \emph{\bibinfo{journal}{npj Quantum Inf.}}  (\bibinfo{year}{2026}).

\bibitem{Ra2020}
\bibinfo{author}{Ra, Y.-S.} \emph{et~al.}
\newblock \bibinfo{title}{{Non-Gaussian quantum states of a multimode light
  field}}.
\newblock \emph{\bibinfo{journal}{Nat. Phys.}} \textbf{\bibinfo{volume}{16}},
  \bibinfo{pages}{144--147} (\bibinfo{year}{2020}).

\bibitem{Kouadou2023}
\bibinfo{author}{Kouadou, T.} \emph{et~al.}
\newblock \bibinfo{title}{{Spectrally shaped and pulse-by-pulse multiplexed
  multimode squeezed states of light}}.
\newblock \emph{\bibinfo{journal}{APL Photonics}} \textbf{\bibinfo{volume}{8}},
  \bibinfo{pages}{086113} (\bibinfo{year}{2023}).

\bibitem{Roh2025}
\bibinfo{author}{Roh, C.}, \bibinfo{author}{Gwak, G.}, \bibinfo{author}{Yoon,
  Y.-D.} \& \bibinfo{author}{Ra, Y.-S.}
\newblock \bibinfo{title}{{Generation of three-dimensional cluster entangled
  state}}.
\newblock \emph{\bibinfo{journal}{Nat. Photon.}} \textbf{\bibinfo{volume}{19}},
  \bibinfo{pages}{526--532} (\bibinfo{year}{2025}).

\bibitem{Kalash2025}
\bibinfo{author}{Kalash, M.}, \bibinfo{author}{Han, U.-N.},
  \bibinfo{author}{Ra, Y.-S.} \& \bibinfo{author}{Chekhova, M.~V.}
\newblock \bibinfo{title}{{Efficient detection of spectrally multimode squeezed
  light through optical parametric amplification}} (\bibinfo{year}{2025}).
\newblock \bibinfo{note}{Preprint at \url{https://arxiv.org/abs/2508.04502}}.

\bibitem{Gwak2026}
\bibinfo{author}{Gwak, G.}, \bibinfo{author}{Roh, C.}, \bibinfo{author}{Yoon,
  Y.-D.}, \bibinfo{author}{Kim, M.~S.} \& \bibinfo{author}{Ra, Y.-S.}
\newblock \bibinfo{title}{{Completely characterizing multimode second-order
  nonlinear optical quantum processes}}.
\newblock \emph{\bibinfo{journal}{Nat. Photon.}} \textbf{\bibinfo{volume}{20}},
  \bibinfo{pages}{156--162} (\bibinfo{year}{2026}).

\bibitem{Roh2026}
\bibinfo{author}{Roh, C.}, \bibinfo{author}{Gwak, G.}, \bibinfo{author}{Yoon,
  Y.-D.} \& \bibinfo{author}{Ra, Y.-S.}
\newblock \bibinfo{title}{{Experimental Quantum Tomography of Multimode
  Gaussian States}} (\bibinfo{year}{2026}).
\newblock \bibinfo{note}{Preprint at \url{https://arxiv.org/abs/2603.21380}}.

\bibitem{Bimbard2010}
\bibinfo{author}{Bimbard, E.}, \bibinfo{author}{Jain, N.},
  \bibinfo{author}{MacRae, A.} \& \bibinfo{author}{Lvovsky, A.~I.}
\newblock \bibinfo{title}{{Quantum-optical state engineering up to the
  two-photon level}}.
\newblock \emph{\bibinfo{journal}{Nat. Photon.}} \textbf{\bibinfo{volume}{4}},
  \bibinfo{pages}{243--247} (\bibinfo{year}{2010}).

\bibitem{Endo2023}
\bibinfo{author}{Endo, M.} \emph{et~al.}
\newblock \bibinfo{title}{{Non-Gaussian quantum state generation by
  multi-photon subtraction at the telecommunication wavelength}}.
\newblock \emph{\bibinfo{journal}{Opt. Express}} \textbf{\bibinfo{volume}{31}},
  \bibinfo{pages}{12865--12879} (\bibinfo{year}{2023}).

\bibitem{Simon2024}
\bibinfo{author}{Simon, H.}, \bibinfo{author}{Caron, L.},
  \bibinfo{author}{Journet, R.}, \bibinfo{author}{Cotte, V.} \&
  \bibinfo{author}{Tualle-Brouri, R.}
\newblock \bibinfo{title}{{Experimental Demonstration of a Versatile and
  Scalable Scheme for Iterative Generation of Non-Gaussian States of Light}}.
\newblock \emph{\bibinfo{journal}{Phys. Rev. Lett.}}
  \textbf{\bibinfo{volume}{133}}, \bibinfo{pages}{173603}
  (\bibinfo{year}{2024}).

\bibitem{Barakat2025}
\bibinfo{author}{Barakat, I.} \emph{et~al.}
\newblock \bibinfo{title}{{Simultaneous measurement of multimode squeezing
  through multimode phase-sensitive amplification}}.
\newblock \emph{\bibinfo{journal}{Opt. Quantum}} \textbf{\bibinfo{volume}{3}},
  \bibinfo{pages}{36--44} (\bibinfo{year}{2025}).

\bibitem{Lamine2008}
\bibinfo{author}{Lamine, B.}, \bibinfo{author}{Fabre, C.} \&
  \bibinfo{author}{Treps, N.}
\newblock \bibinfo{title}{{Quantum Improvement of Time Transfer between Remote
  Clocks}}.
\newblock \emph{\bibinfo{journal}{Phys. Rev. Lett.}}
  \textbf{\bibinfo{volume}{101}}, \bibinfo{pages}{123601}
  (\bibinfo{year}{2008}).

\bibitem{Wang2018}
\bibinfo{author}{Wang, S.} \emph{et~al.}
\newblock \bibinfo{title}{{Sub-shot-noise interferometric timing measurement
  with a squeezed frequency comb}}.
\newblock \emph{\bibinfo{journal}{Phys. Rev. A}} \textbf{\bibinfo{volume}{98}},
  \bibinfo{pages}{053821} (\bibinfo{year}{2018}).

\bibitem{Bachor2019}
\bibinfo{author}{Bachor, H.-A.} \& \bibinfo{author}{Ralph, T.~C.}
\newblock \emph{\bibinfo{title}{{A Guide to Experiments in Quantum Optics}}}
  (\bibinfo{publisher}{Wiley-VCH}, \bibinfo{address}{Weinheim},
  \bibinfo{year}{2019}), \bibinfo{edition}{3} edn.

\bibitem{Kumar2012}
\bibinfo{author}{Kumar, R.} \emph{et~al.}
\newblock \bibinfo{title}{{Versatile wideband balanced detector for quantum
  optical homodyne tomography}}.
\newblock \emph{\bibinfo{journal}{Opt. Commun.}}
  \textbf{\bibinfo{volume}{285}}, \bibinfo{pages}{5259--5267}
  (\bibinfo{year}{2012}).

\bibitem{Cooper2013}
\bibinfo{author}{Cooper, M.}, \bibinfo{author}{S{\"o}ller, C.} \&
  \bibinfo{author}{Smith, B.~J.}
\newblock \bibinfo{title}{{High-stability time-domain balanced homodyne
  detector for ultrafast optical pulse applications}}.
\newblock \emph{\bibinfo{journal}{J. Mod. Opt.}} \textbf{\bibinfo{volume}{60}},
  \bibinfo{pages}{611--616} (\bibinfo{year}{2013}).

\bibitem{Masalov2017}
\bibinfo{author}{Masalov, A.~V.}, \bibinfo{author}{Kuzhamuratov, A.} \&
  \bibinfo{author}{Lvovsky, A.~I.}
\newblock \bibinfo{title}{{Noise spectra in balanced optical detectors based on
  transimpedance amplifiers}}.
\newblock \emph{\bibinfo{journal}{Rev. Sci. Instrum.}}
  \textbf{\bibinfo{volume}{88}}, \bibinfo{pages}{113109}
  (\bibinfo{year}{2017}).

\bibitem{Hansen2001}
\bibinfo{author}{Hansen, H.} \emph{et~al.}
\newblock \bibinfo{title}{{Ultrasensitive pulsed, balanced homodyne detector:
  application to time-domain quantum measurements}}.
\newblock \emph{\bibinfo{journal}{Opt. Lett.}} \textbf{\bibinfo{volume}{26}},
  \bibinfo{pages}{1714--1716} (\bibinfo{year}{2001}).

\bibitem{Zavatta2006}
\bibinfo{author}{Zavatta, A.}, \bibinfo{author}{Viciani, S.} \&
  \bibinfo{author}{Bellini, M.}
\newblock \bibinfo{title}{{Non-classical field characterization by
  high-frequency, time-domain quantum homodyne tomography}}.
\newblock \emph{\bibinfo{journal}{Laser Phys. Lett.}}
  \textbf{\bibinfo{volume}{3}}, \bibinfo{pages}{3--16} (\bibinfo{year}{2006}).

\bibitem{Okubo2008}
\bibinfo{author}{Okubo, R.}, \bibinfo{author}{Hirano, M.},
  \bibinfo{author}{Zhang, Y.} \& \bibinfo{author}{Hirano, T.}
\newblock \bibinfo{title}{{Pulse-resolved measurement of quadrature phase
  amplitudes of squeezed pulse trains at a repetition rate of 76 MHz}}.
\newblock \emph{\bibinfo{journal}{Opt. Lett.}} \textbf{\bibinfo{volume}{33}},
  \bibinfo{pages}{1458--1460} (\bibinfo{year}{2008}).

\bibitem{Du2018}
\bibinfo{author}{Du, S.}, \bibinfo{author}{Li, Z.}, \bibinfo{author}{Liu, W.},
  \bibinfo{author}{Wang, X.} \& \bibinfo{author}{Li, Y.}
\newblock \bibinfo{title}{{High-speed time-domain balanced homodyne detector
  for nanosecond optical field applications}}.
\newblock \emph{\bibinfo{journal}{J. Opt. Soc. Am. B}}
  \textbf{\bibinfo{volume}{35}}, \bibinfo{pages}{481--486}
  (\bibinfo{year}{2018}).

\bibitem{Kouadou2026}
\bibinfo{author}{Kouadou, T.} \emph{et~al.}
\newblock \bibinfo{title}{{Homodyne detection for pulse-by-pulse squeezing
  measurements}}.
\newblock \emph{\bibinfo{journal}{New J. Phys.}} \textbf{\bibinfo{volume}{28}},
  \bibinfo{pages}{054509} (\bibinfo{year}{2026}).

\bibitem{Appel2007}
\bibinfo{author}{Appel, J.}, \bibinfo{author}{Hoffman, D.},
  \bibinfo{author}{Figueroa, E.} \& \bibinfo{author}{Lvovsky, A.~I.}
\newblock \bibinfo{title}{{Electronic noise in optical homodyne tomography}}.
\newblock \emph{\bibinfo{journal}{Phys. Rev. A}} \textbf{\bibinfo{volume}{75}},
  \bibinfo{pages}{035802} (\bibinfo{year}{2007}).

\bibitem{Vahlbruch2016}
\bibinfo{author}{Vahlbruch, H.}, \bibinfo{author}{Mehmet, M.},
  \bibinfo{author}{Danzmann, K.} \& \bibinfo{author}{Schnabel, R.}
\newblock \bibinfo{title}{{Detection of 15 dB Squeezed States of Light and
  Their Application for the Absolute Calibration of Photoelectric Quantum
  Efficiency}}.
\newblock \emph{\bibinfo{journal}{Phys. Rev. Lett.}}
  \textbf{\bibinfo{volume}{117}}, \bibinfo{pages}{110801}
  (\bibinfo{year}{2016}).

\bibitem{Mari2012}
\bibinfo{author}{Mari, A.} \& \bibinfo{author}{Eisert, J.}
\newblock \bibinfo{title}{{Positive Wigner Functions Render Classical
  Simulation of Quantum Computation Efficient}}.
\newblock \emph{\bibinfo{journal}{Phys. Rev. Lett.}}
  \textbf{\bibinfo{volume}{109}}, \bibinfo{pages}{230503}
  (\bibinfo{year}{2012}).

\bibitem{Aasi2013}
\bibinfo{author}{Aasi, J.} \emph{et~al.}
\newblock \bibinfo{title}{{Enhanced sensitivity of the LIGO gravitational wave
  detector by using squeezed states of light}}.
\newblock \emph{\bibinfo{journal}{Nat. Photon.}} \textbf{\bibinfo{volume}{7}},
  \bibinfo{pages}{613--619} (\bibinfo{year}{2013}).

\bibitem{Gessner2019}
\bibinfo{author}{Gessner, M.}, \bibinfo{author}{Smerzi, A.} \&
  \bibinfo{author}{Pezz{\'e}, L.}
\newblock \bibinfo{title}{{Metrological Nonlinear Squeezing Parameter}}.
\newblock \emph{\bibinfo{journal}{Phys. Rev. Lett.}}
  \textbf{\bibinfo{volume}{122}}, \bibinfo{pages}{090503}
  (\bibinfo{year}{2019}).

\bibitem{Go2026}
\bibinfo{author}{Go, B.-Y.} \emph{et~al.}
\newblock \bibinfo{title}{{Quantum metrology under coarse-grained
  measurement}}.
\newblock \emph{\bibinfo{journal}{Opt. Express}} \textbf{\bibinfo{volume}{34}},
  \bibinfo{pages}{17346--17359} (\bibinfo{year}{2026}).

\bibitem{Gambetta2007}
\bibinfo{author}{Gambetta, J.}, \bibinfo{author}{Braff, W.~A.},
  \bibinfo{author}{Wallraff, A.}, \bibinfo{author}{Girvin, S.~M.} \&
  \bibinfo{author}{Schoelkopf, R.~J.}
\newblock \bibinfo{title}{{Protocols for optimal readout of qubits using a
  continuous quantum nondemolition measurement}}.
\newblock \emph{\bibinfo{journal}{Phys. Rev. A}} \textbf{\bibinfo{volume}{76}},
  \bibinfo{pages}{012325} (\bibinfo{year}{2007}).

\bibitem{Eaton2023}
\bibinfo{author}{Eaton, M.} \emph{et~al.}
\newblock \bibinfo{title}{{Resolution of 100 photons and quantum generation of
  unbiased random numbers}}.
\newblock \emph{\bibinfo{journal}{Nat. Photon.}} \textbf{\bibinfo{volume}{17}},
  \bibinfo{pages}{106--111} (\bibinfo{year}{2023}).

\end{thebibliography}
\providecommand{\noopsort}[1]{}\providecommand{\singleletter}[1]{#1}%

\end{document}